\documentclass[twocolumn]{aastex6}

\usepackage{graphicx}
\usepackage{amssymb}
\usepackage{amsmath}
\usepackage{layout}
\usepackage{url}
\usepackage{subfigure}
\usepackage{color}
\usepackage{enumerate}
\newcommand\T{\rule{0pt}{2.6ex}}       
\newcommand\B{\rule[-1.2ex]{0pt}{0pt}} 

\def\beq{\begin{equation}}
\def\eeq{\end{equation}}
\def\bey{\begin{eqnarray}}
\def\eey{\end{eqnarray}}

\begin{document}

\title{Variability and Optical Polarization Can Probe the Neutrino and Electromagnetic Emission Mechanisms of TXS~0506+056}


\author{Haocheng Zhang\altaffilmark{1}, Ke Fang\altaffilmark{2}, Hui Li\altaffilmark{3}}

\altaffiltext{1}{Department of Physics and Astronomy, Purdue University, West Lafayette, IN 47907, USA}

\altaffiltext{2}{JSI Fellow, Department of Astronomy, University of Maryland, College Park, MD 20742, USA}

\altaffiltext{3}{Theoretical Division, Los Alamos National Laboratory, Los Alamos, NM 87545, USA}

\begin{abstract}

The association of the high-energy neutrino event IceCube-170922A with the flaring blazar TXS~0506+056 indicates that hadronic processes may operate in a blazar jet. We perform semi-analytical spectral fitting of the multi-wavelength emission to obtain estimates of the jet physical parameters, and find that the multi-wavelength emission can be explained by either a proton synchrotron scenario or an electron inverse Compton scattering scenario. In the first scenario, the high-energy component of the spectral energy distribution is dominantly contributed by the synchrotron emission of primary protons. A strong magnetic field of $10-100$~G is required, implying that the particle acceleration is likely driven by magnetic energy dissipation such as magnetic reconnection events. In the second scenario, the inverse Compton emission of primary electrons induces the high-energy component, implying a magnetic field of $0.1-1$~G. Thus the particle acceleration is likely driven by the kinetic energy dissipation such as shocks. We also discuss the neutrino production in the context of single-zone and multi-zone models based on the above two scenarios.
We demonstrate that the variability and optical polarization signatures can be used to distinguish the two scenarios due to their drastically different magnetic field. Specifically, the proton synchrotron scenario may show orphan fast variability in the low-energy spectral component on top of the active state, with an optical polarization degree $\lesssim 10\%$ throughout the active state. The inverse Compton scattering scenario instead predicts co-variability of the low- and high-energy components on both short and long time scales, as well as a strongly variable optical polarization degree that can reach $\gtrsim 20\%$. Our results suggest that optical polarization measurements and well-resolved multi-wavelength light curves can be used to understand the electromagnetic and high-energy neutrino emissions by TXS~0506+056 and similar events in the future.

\end{abstract}


\section{Introduction}

Since operation, IceCube neutrino observatory has opened up a new window to observe the universe \citep[see e.g.,][]{IceCube13}. Neutrinos are generally associated with hadronic processes in the universe, thus they may point to the origin of cosmic rays. Extragalactic sources, such as active galactic nuclei \citep{Mannheim92b,Dermer14,Tavecchio15} and starburst galaxies \citep{Murase13,Liu14,Tamborra14}, have been suggested as the potential high energy neutrino sources. 

The IceCube-170922A event, a  $\sim 290$~TeV neutrino, was recently reported to be coincident with the blazar TXS~0506+056 during its flaring state
\citep{IceCube18a,IceCube18b}. Following the neutrino alert, the blazar was detected in multi-wavelength campaigns including very-high-energy $\gamma$-rays \citep{IceCube18a, MAGIC18, 2018ApJ...861L..20A}. All wavelengths exhibit strong variability, with a $\sim 7\%$ optical polarization degree during flares reported by the Kanata Telescope \citep{IceCube18a}. Analysis of historical IceCube data independently found a $3.5\,\sigma$ excess of neutrinos from the direction of this blazar \citep{IceCube18b}. Follow-up searches by the ANTARES telescope (10 GeV to 100 TeV) reported a consistency with the background \citep{2018arXiv180704309A}. A dissection of the region around the neutrino further suggests that TXS~0506+056 could be a high-energy neutrino source \citep{2018arXiv180704461P}.

 
Blazars are active galactic nuclei whose jet is pointing very close to our line of sight. They exhibit highly variable emission from radio up to TeV $\gamma$-rays \citep{Ackermann16}. Their spectral energy distribution (SED) are typically composed of two parts: a low-energy component from radio up to soft X-ray and a high-energy component from X-ray up to TeV $\gamma$-ray. The low-energy component is generally believed to be the synchrotron emission of primary electrons, as it usually exhibits a high degree of polarization \citep{Scarpa97}. In particular, the optical polarization signatures can be highly variable alongside multi-wavelength flares \citep{Angelakis16}. The origin of the high-energy component may be either leptonic or hadronic \citep{Boettcher13,Cerruti15}. In a leptonic scenario, the high-energy emission is produced by primary electrons upper scattering background photons through the inverse Compton (IC) process. The background photons can be the low-energy synchrotron emission by the same population of electrons \citep[synchrotron-self Compton, SSC, e.g.,][]{Marscher85,Maraschi92} or external photon fields such as the thermal radiation by the accretion disk, the broad line region (BLR), and the dusty torus \citep[external Compton, EC, e.g.,][]{Dermer92,Sikora94}. In a hadronic scenario, the high-energy component is produced by the primary proton synchrotron (PS) or the synchrotron of secondary charged particles from hadronic interactions \citep{Mannheim92a,Mucke01,Petropoulou15}. High-energy neutrinos, which come from the decay of charged pions, are a unique signature of such hadronic interactions. In addition, the PS scenario predicts a high polarization degree in the X-ray and $\gamma$-ray bands \citep{Zhang13,Paliya18}, which may be probed by future high-energy polarimeters such as IXPE and AMEGO \citep{Weisskopf16,McEnery17}.


Theoretical models have been proposed to explain the SED of TXS~0506+056 and the presence of IceCube-170922A. These models may be categorized into two generic groups according to the domination of hadronic processes in the high-energy spectral component. The first type of models mainly attribute the high-energy component to the IC emission of primary electrons \citep[e.g.,][]{MAGIC18, Cerruti18, Gao18, Keivani18}, while the hadronic interactions only play a minor role in the multi-wavelength SED.
The second type of models argue that the high-energy spectral component has a significant contribution from hadronic processes, specifically by the PS \citep{Cerruti18} or by pion emissions and electromagnetic cascades \citep{Murase18,Liu18}.


In this paper, we focus on the variability and polarization signatures that arise from the drastically different physical conditions implied by the PS and IC scenarios. We demonstrate that these generic observable signatures can provide further constraints to distinguish theoretical models. In section 2, we estimate the parent proton energy and the target photon energy for neutrino production through photopion process, and show that a pure hadronic model to explain the whole SED is disfavored. Section 3 performs multi-wavelength spectral fitting and neutrino flux estimates based on leptohadronic models, and discusses the underlying particle acceleration processes. Section 4 illustrates the physical constraints that can be derived from multi-wavelength variability and optical polarization. Finally we summarize our results in Section 5.

\section{General Estimates}

We assume that the blazar jet is relativistic with a bulk Lorentz factor $\Gamma= 10\,\Gamma_1$, so that the  Doppler factor at a viewing angle $\theta_{\rm obs}$ from the jet axis is $\delta\equiv [\Gamma(1-\beta_{\Gamma}\cos \theta_{obs})]^{-1}= 10\,\delta_1$. In the following context all quantities in the comoving frame of the blazar jet are marked with a prime. The comoving size of the emission region may be inferred by the causality relation
\beq
R' \lesssim \delta c t_{\rm var} / (1+z) \sim 10^{17}\,\delta_1\,t_{\rm var, 6}\,\rm cm~~, 
\eeq 
where $t_{\rm var}=10^6\,t_{\rm var, 6}\,\rm s$ is the variability time scale. We adopt a redshift $z = 0.3365$ \citep{Paiano18} for TXS~0506-056 throughout the work. The IceCube neutrino with an observed energy of $E_{\nu}\sim 290~\rm{TeV}$ \citep{IceCube18a} indicates that the neutrino energy in the comoving frame is 
\beq
E_\nu' = \frac{E_\nu (1+z)}{\delta} \sim 40 \,\delta_1\,\rm TeV~~,
\eeq
and the required energy of the parent proton is
\beq
E_p' \approx 20\,E_\nu' \sim 800\,\delta_1\,\rm TeV~~, 
\eeq
or equivalently
\beq
\gamma_p' \sim 8\times10^5\,\delta_1~~. 
\eeq
For a $\sim$PeV proton that produced the observed muon neutrino, the target photon energy normalized by the electron rest energy that corresponds to the $\Delta$-resonance of the photopion production is 
\beq
\epsilon'_{p\gamma} \approx \frac{\bar{\epsilon}_\Delta}{2\gamma_p'm_ec^2} = 3\times 10^{-4}\,\gamma_{p,6}'^{-1}~~,
\eeq
where we put $\gamma'_p=10^6 \gamma'_{p,6}$ and $\bar{\epsilon}_\Delta\sim 0.3$~GeV is the resonance energy (e.g.\citealp{2006PhRvD..73f3002M}).


A pure hadronic model, where the multi-wavelength emission exclusively originates from the primary protons and their secondary products, cannot explain the observed TXS~0506+056 spectrum. The PS emission clearly cannot explain the low-energy spectral component. The synchrotron critical frequency is \citep{Rybicki79}
\begin{equation}
\nu'_{syn}=\frac{3eB'}{4\pi mc}\gamma'^2\sim 4\times 10^6 \frac{m_e}{m}B'\gamma'^2~\rm{Hz}~~.
\label{syn}
\end{equation}
If this corresponds to the low-energy spectral peak at the optical band $\nu_{low}\sim 10^{15}\nu_{low,15}~\rm{Hz}$, the magnetic field in the comoving frame is then $B'\sim 0.05\delta_1^{-1}\gamma^{\prime -2}_{p,6}\nu_{low,15}~\rm{G}$. Obviously the PS mechanism is not efficient with such a low magnetic field.


If the high-energy spectral component is dominated by the PS radiation, which generally requires tens of Gauss of magnetic field, then the low-energy spectral component cannot be explained by the secondary hadronic products either. The neutral and charged pion decays give photons and secondary electrons at $\sim 40\delta_1~\rm{TeV}$ similar to the neutrinos. The synchrotron emission or the Compton scattering of cosmic microwave background (CMB) photons by these highly energetic secondary electrons should peak at $\sim 50~\rm{GeV}$ and $\sim 500~\rm{TeV}$, respectively (see Equation (\ref{syn}) and Section~\ref{sec:IC} for SSC of primary electrons).

\section{Lepto-hadronic Models}
\begin{figure}
\begin{center}
\includegraphics[width=.49\textwidth]{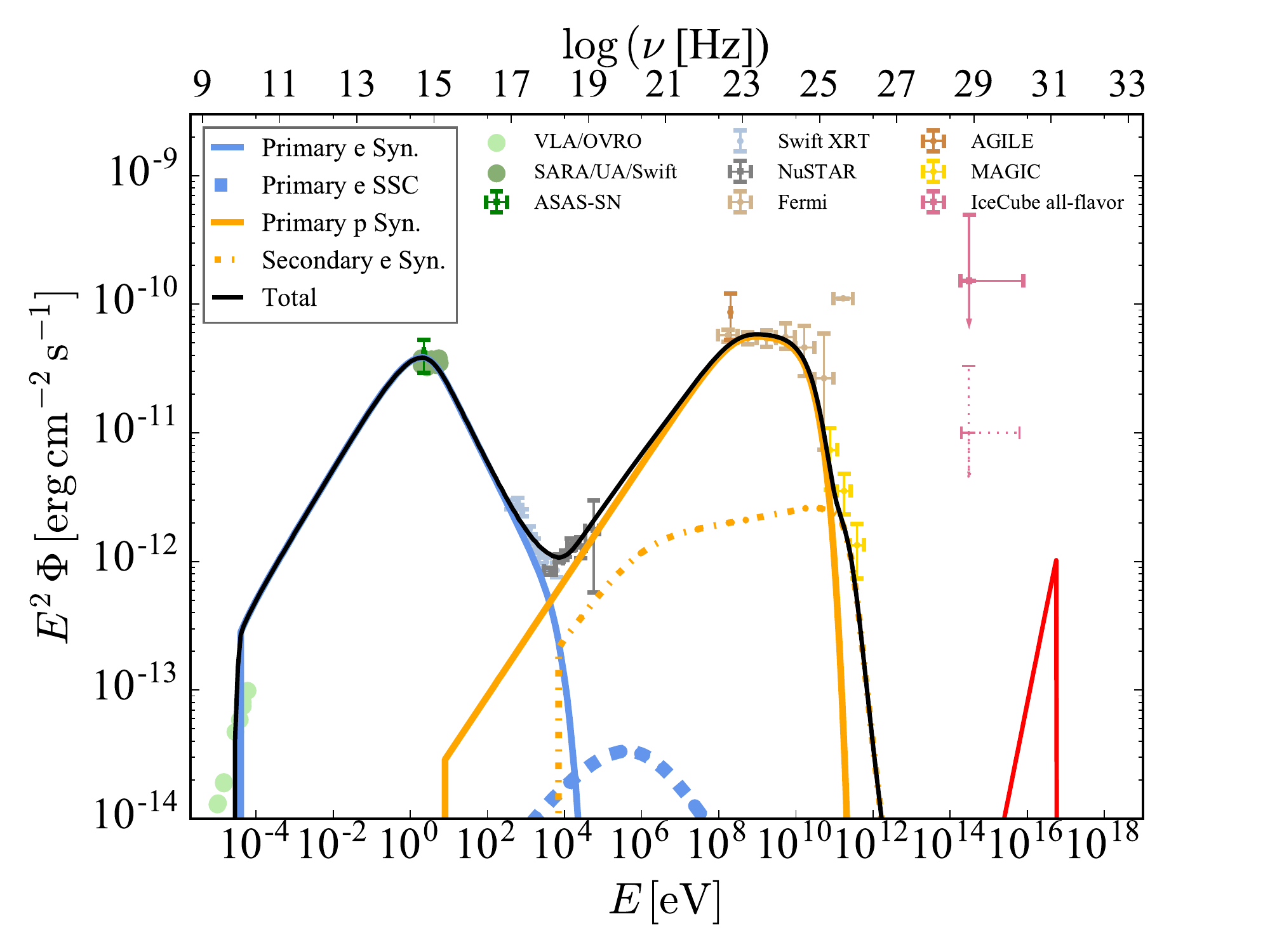}
\includegraphics[width=.49\textwidth]{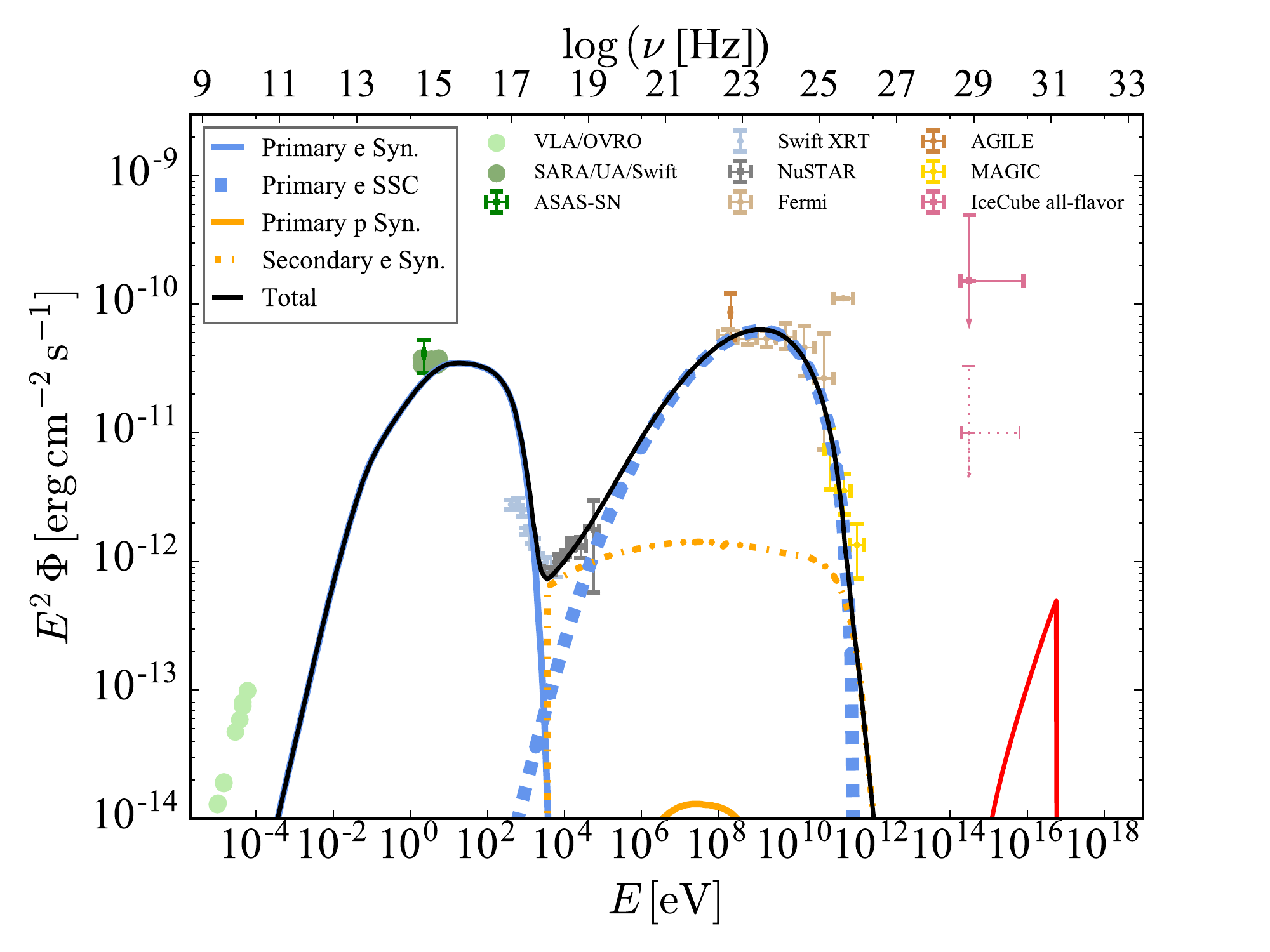}
\end{center}
\caption{\label{fig:default} Multi-wavelength spectral fitting using the proton synchrotron (PS, upper panel) scenario and the inverse Compton (IC, lower panel) scenario comparing to the TXS~0506+056 observation \citep{IceCube18a}. The model parameters are listed in Table~\ref{table}. In both plots, the blue curves represent the synchrotron emission (solid) and inverse Compton emission (thick dashed) by primary electrons, orange curves correspond to photons emitted by primary protons (solid) and their secondaries (dash-dotted), and red solid curves show neutrinos from the photopion production of primary protons.   
}
\end{figure}

In this section we discuss lepto-hadronic models that can explain the SED of TXS~0506+056.  While the low-energy spectral component is dominated by the primary electron synchrotron, we show that the high-energy component may be either dominated by the PS mechanism (Section~\ref{sec:PS}) or by the IC mechanism (Section~\ref{sec:IC}). Using both general arguments derived from observed spectral features and semi-analytical spectral fitting with default parameters, we demonstrate that drastically different magnetic fields and particle acceleration mechanisms are predicted by the two scenarios. Our spectral fitting is done with a stationary lepto-hadronic radiation code developed by \cite{Boettcher13}, which semi-analytically treats radiative, photomeson, and adiabatic cooling as well as particle escaping. The default particle escaping time scale is chosen as 4 times of the light crossing time scale. The derived SED is corrected by EBL attenuation. Figure \ref{fig:default} shows our fitting results with default parameters listed in Table \ref{table}. Estimates of the neutrino flux in the context of one-zone and multi-zone models are done analytically in Section \ref{sec:fpgamma}.

\begin{table*}[ht]
\scriptsize
\parbox{\linewidth}{
\centering
\begin{tabular}{|l|c|c|}\hline
\multicolumn{3}{|c|}{Model Parameters}                                                    \\ \hline
                                              & Proton Synchrotron & Inverse Compton      \\ \hline
Redshift ($z$)                                & \multicolumn{2}{|c|}{0.3365}              \\ \hline
Bulk Lorentz factor ($\Gamma$)                & \multicolumn{2}{|c|}{10}                  \\ \hline
Viewing angle ($\theta_{obs}$)                & \multicolumn{2}{|c|}{0}                   \\ \hline
Blob size ($R'$, cm)                          & \multicolumn{2}{|c|}{$5\times 10^{15}$}   \\ \hline
Escape time ($t'_{esc}$, s)                   & \multicolumn{2}{|c|}{$6.67\times 10^{5}$} \\ \hline
Magnetic field ($B'$, G)                      & 50                 & 1.2                  \\ \hline
Minimal electron Lorentz factor ($\gamma'_{e,min}$) & 400          & 4000                 \\ \hline
Maximal electron Lorentz factor ($\gamma'_{e,max}$) & 20000        & 40000                \\ \hline
Electron power-law index ($\alpha_e$)         & 3.2                & 2.0                  \\ \hline
Electron kinetic luminosity ($L_{e,kin}$, $\rm{erg\,s^{-1}}$) & $6\times 10^{43}$ & $3.2\times 10^{44}$ \\ \hline
Minimal proton Lorentz factor ($\gamma'_{p,min}$) & \multicolumn{2}{|c|}{1}               \\ \hline
Turnover proton Lorentz factor ($\gamma'_{p,b}$)  & \multicolumn{2}{|c|}{$2\times 10^8$}  \\ \hline
Maximal proton Lorentz factor ($\gamma'_{p,max}$) & \multicolumn{2}{|c|}{$2\times 10^9$}  \\ \hline
Proton power-law index ($\alpha_p$)               & \multicolumn{2}{|c|}{2.1}             \\ \hline
Proton power-law index after turnover ($\beta_p$) & \multicolumn{2}{|c|}{3.0}             \\ \hline
Proton kinetic luminosity ($L_{p,kin}$, $\rm{erg\,s^{-1}}$) & $8\times 10^{46}$ & $3.2\times 10^{46}$ \\ \hline
\multicolumn{3}{|c|}{Derived Quantities}                                                  \\ \hline
Jet Power ($P^{obs}_{jet}=L_{e,kin}+L_{p,kin}+L_B$, $\rm{erg\,s^{-1}}$) & $1.04\times 10^{47}$ & $3.2\times 10^{46}$ \\ \hline
$L_B/L_{p,kin}$                   & 0.3                & 0.0004               \\ \hline
$L_B/L_{e,kin}$                   & 400                & 0.04                 \\ \hline
$L_{p,kin}/L_{e,kin}$             & 1000               & 100                  \\ \hline
\end{tabular}}
\caption{Parameters and derived quantities of the default proton synchrotron and inverse Compton models shown in Figure~\ref{fig:default}. \label{table}}
\end{table*}

\subsection{Proton Synchrotron Scenario}\label{sec:PS}

In the PS scenario the proton synchrotron dominates the entire high-energy spectral component, while SSC from primary electrons and secondary electron synchrotron make trivial contributions. This typically requires a very high magnetic field of $10-100~\rm{G}$ \citep{Boettcher13}. Using a default magnetic field of $B'_{PS}=50~\rm{G}$, Equation (\ref{syn}) shows that to obtain the {\it Fermi} $\gamma$-rays, the primary proton Lorentz factor should extend to $\gamma'_p \sim 10^9$. The proton synchrotron cooling rate is 
\begin{equation}
\dot{\gamma}'_{syn}=-\frac{4}{3}c\sigma_T\frac{u'_B}{m_e c^2}(\frac{m_e}{m_p})^3 \gamma'^2\sim -10^{-9}(\frac{m_e}{m})^3B^{\prime 2}\gamma'^2~~,
\end{equation}
where $\sigma_T$ is the Thomson cross section and $u'_B = B^{\prime 2}/8\pi$ is the magnetic energy density. This corresponds to a cooling time of 
\beq
t_{\rm PS}' = \frac{\gamma_p'}{|\dot{\gamma}_{\rm PS}'|} = 2\times10^6\, \delta_1^2 \,\gamma_{p,9}'^3\,\rm s. 
\eeq
The flux of the synchrotron emission may be approximated as \citep{DermerBook}
\beq\label{eqn:flux_PS}
\left(\epsilon F_\epsilon \right)^{\rm obs}_{\rm PS}\approx\frac{\delta^4}{8\pi d_L^2}\,\frac{V_b' \gamma_s'^2 n'({\gamma_s'}) m_p c^2}{ t_{\rm PS}'(\gamma_s')}
\eeq
with 
\beq
\gamma_s' = \left(\frac{(1+z)}{\delta}\frac{\epsilon}{\epsilon_B'}\right)^{1/2},
\eeq
where $\epsilon_B' \equiv 2h\pi m_p c / eB'$ and  $V_b' = 4\pi R'^3/3$ is the volume of the relativistic blob.

Assuming that protons follow a simple power law up to a break energy $\gamma_{p,\rm b}'$, 
\beq\label{eqn:n_gammap}
n'(\gamma_p') = k_p'\gamma_p'^{-s}\, \quad\quad 1< \gamma_p' < \gamma_{p, \rm b}'. 
\eeq
then the flux of the synchrotron emission follows 
\beq
\left(\epsilon F_\epsilon \right)^{\rm obs}_{\rm PS} \propto \epsilon^{(3-s) / 2}
\eeq
up to the peak frequency. 
The high-energy spectral component follows $\sim \epsilon^{0.4}$ from 10~keV to GeV and has a peak flux of $\left(\epsilon F_\epsilon\right)_{\rm pk2}^{\rm obs} \sim 6\times10^{-11}\,\rm erg\,cm^{-2}\,s^{-1}$,  implying that protons follow a spectral index of $s \sim 2.2$ and an energy density 
\bey\label{eqn:u_p'_PS}
u_p' &=& \int_1^{\gamma_{p,\rm b}'} d\gamma_p' \,n'({\gamma_p'})\, \gamma_p'\, m_p c^2  \\ \nonumber
&=& 0.07\,\delta_1^{-5}t_{\rm var,6}^{-3}\,\gamma_{p,\rm b,9}'^{(1+s)}\,\rm erg\,cm^{-3}. 
\eey
The energy density corresponds to an absolute proton luminosity after accounting for the angular distribution of the radiative power \citep{2012ApJ...755..147D} 
\bey\label{eqn:Lp_PS}
L_{p,\rm abs} &\approx& \frac{8\Gamma^2}{3\delta^4}L_p = \frac{8\Gamma^2}{3} 4\pi  R'^2 c  u_p' \\ \nonumber 
&=&  3.4\times10^{47}\,\delta_1^{-1}t_{\rm var,6}^{-1}\,\gamma_{p,\rm b, 9}'^{1+s}\,\rm erg\,s^{-1}. 
\eey
As shown in Figure \ref{fig:default} and Table \ref{table}, our fitting parameters are consistent with the above order-of-magnitude estimates, and produce reasonable fitting to the multi-wavelength spectrum.

The high magnetic field in the PS scenario suggests that the jet energy dissipation is likely driven by kink instabilities, and the nonthermal particles are accelerated through magnetic reconnection events \citep{Mizuno09,Guan14,Zhang16a}. Magnetic reconnection is likely to occur during the nonlinear kink stages and numerical simulations have shown that magnetic reconnection can efficiently accelerate electrons and protons into a power-law shape with a spectral turnover within one decade of the maximal particle energy \citep{Guo16,Werner18}. Although the nonthermal electrons are efficiently cooled through synchrotron due to high magnetic field, which is consistent with the observed soft X-ray spectral shape, the proton synchrotron cooling is however inefficient. Thus the $\gamma$-rays should represent the intrinsic proton spectrum, including the high-energy turnover that is seen as the flat {\it Fermi} spectrum at GeV energies. Our model suggests a proton power-law index of $-2.1$ based on the hard X-ray to $\gamma$-ray data. Additionally, our PS scenario suggests that the nonthermal proton energy dominates over the total particle kinetic energy, but is comparable to the jet magnetic power. This is consistent with previous fittings of generic blazar spectra \citep{Boettcher13}. Notice that the radio emission is likely coming from a much larger region than the multi-wavelength flaring region used in our spectral fitting, thus we observe that our spectral fitting quickly cuts off at radio frequencies.

\subsection{Inverse Compton Scenario}\label{sec:IC}

In the IC scenario, we suggest that the high-energy component is dominated by primary electron SSC. The SSC photon frequency is then given by
\begin{equation}
\nu'_{SSC}=\nu'_{syn}\gamma^{\prime 2}_e
\end{equation}
Given that the low-energy spectral component peaks at optical and the high-energy component at GeV $\gamma$-ray, one can easily find $\gamma'_e\sim 3\times 10^4\gamma'_{e,4.5}$. Thus the magnetic field strength in the IC scenario can be obtained by Equation (\ref{syn}), $B'_{IC}\sim 1 \delta_1^{-1}\gamma_{e,4.5}'^{-2}\,\nu_{\rm pk1, 16}~\rm G$.

The ratio $\cal{R}$ of the peak fluxes of the SSC and the synchrotron emissions from a power-law electron distribution, $n'(\gamma_e') = k_e' \,\gamma_e'^{-s}$, can be written as \citep{DermerBook}
\beq
{\cal R} = \frac{\left(\epsilon F_\epsilon\right)_{\rm SSC,pk} }{\left(\epsilon F_\epsilon\right)_{\rm syn,pk}} \approx \frac{2}{3} \sigma_T R' \Sigma_c k_e' \gamma_{e,b}'^{3-s} 
\eeq
with $\Sigma_c$ being a factor of the order unity \citep{DermerBook}. 
The low- and high-energy components of the SED of TXS~0506+056 present comparable peak fluxes, suggesting that the electron density is 
\bey
u_e' &=& \int_1^{\gamma_{e,b}'} d\gamma_e'\gamma_e' m_e c^2 n'(\gamma_e') \\ \nonumber
&=& 3\times10^{-3}\,{\cal R}\delta_1^{-1}t_{\rm var,6}^{-1} \Sigma_c\gamma_{e,\rm b,4.5}'^{-1}\,\rm erg\,cm^{-3}, 
\eey
if taking $s\sim 2$ as required by the hard X-ray and $\gamma$-ray data. 
The energy density corresponds to an absolute electron luminosity 
\beq
L_{e,\rm abs} \approx \frac{8 \Gamma^2}{3} 4\pi R'^2 c u_e' \approx 5\times10^{45}\,\delta_1^{4}t_{\rm var,6}^{2}\,u_{e,-3}'\,\rm erg\,s^{-1}
\eeq


The IC scenario cannot, however, constrain the proton kinetic luminosity. Here we assume conventionally that the proton energy content is 100 times larger than that of electrons, 
\beq\label{eqn:Lp_IC}
u_p' \sim 100\,u_e'.
\eeq
Using the same power-law of $-2.1$ as in the PS scenario, Figure \ref{fig:default} confirms a trivial contribution from protons for the high-energy spectral component, consistent with the SSC dominance. 

The key difference of IC scenario from the PS scenario is the low magnetic field strength. Under this condition, the jet energy dissipation is likely due to shocks that dissipate the jet bulk kinetic energy. Protons and electrons are then accelerated at the shock front. Also due to the low magnetic field, only high-energy electrons are in the fast cooling regime, thus we observe a cooling spectral break, even though our electron injection spectrum is a single power-law. This explains the spectral turnover in the {\it Fermi} GeV bands.

\subsection{Neutrino Flux \label{sec:fpgamma}}

High-energy neutrinos are produced when primary protons interact with background photons in the jet. Using the measurement of the optical to X-ray flux of TXS~0506+056 we can derive the energy density of the target photons and the effective optical depth for the $p\gamma$ interaction ({\it aka} the pion production efficiency).

If the target photon comes from the emission region itself, its observed energy should be $E^{obs}\sim 1.5\delta_1~\rm{keV}$.
A flux of $\epsilon F_\epsilon \sim 10^{-12}\,\rm erg\,cm^{-2}\,s^{-1}$  is observed at keVs, leading to 
\beq\label{eqn:f_pg_obs}
f'_{p\gamma} \sim\left( \frac{\epsilon\,F_\epsilon\, 4\pi d_L^2 }{\delta^4\,4\pi R'^2 c \,\epsilon'}\right) \sigma_{p\gamma} R' =3\times10^{-7}\,\delta_1^{-5}\,t_{\rm var,6}^{-1}, 
\eeq
where $\sigma_{p\gamma}\approx 1.4\times10^{-28}\,\rm cm^2$ is the effective $p\gamma$ cross section. The same photon field should attenuate the VHE $\gamma$-rays at $\sim 100~\rm{GeV}$. Its cross section for $\gamma\gamma$ absorption is $\sigma_{\gamma\gamma}\sim 10^{-25}~\rm{cm^2}\sim 10^3\sigma_{p\gamma}$ \citep{Boettcher12}, thus the optical depth of $\sim 100~\rm{GeV}$ photons should be $\tau_{\gamma\gamma}\sim 3\times 10^{-4} \ll 1$. Therefore, the neutrinos should be simultaneously produced with the VHE $\gamma$-rays. The fluxes of neutrinos produced by primary protons interacting with the low-energy component in our default cases are shown in Figure~\ref{fig:default}. Our estimates suggest that the neutrino flux obtained within the multi-wavelength emission region itself should be much lower than the observed level, consistent with similar calculations done by \cite{Keivani18} and \cite{Murase18}.

Here we propose a two-zone leptohadronic model. We envision that the relativistic jet, containing  highly accelerated protons and electrons, passes through the broad line region (BLR) and continues to move away from the central black hole, during which the jet dissipates its energy continuously. Here, the two zones refer to the BLR and the region beyond, which changes from being optically thick to optically thin to high energy $\gamma$ rays. The highly energetic protons interact with the dense photon field in the BLR and produce neutrinos through photomeson processes. The $\sim 100~\rm{GeV}$ $\gamma$-rays, however, can hardly escape during the neutrino production phase due to the large optical depth.  As the jet exits out of the BLR, and neutrino production greatly reduces due to the decrease of target photon density. On the other hand, the $\sim 100~\rm{GeV}$ $\gamma$-rays begin to escape. Therefore, this transition through the two zones predicts a delay of the $100~\rm{GeV}$ flare from the neutrino event, but the light curves below $\sim 10~\rm{GeV}$ should appear simultaneous with the neutrino event. Since the BLR is stationary in the host galaxy frame, its UV photon field gets a Lorentz boost to the soft X-ray band in the jet comoving frame. Since we do not observe a UV bump in the multi-wavelength spectrum \citep{IceCube18a}, the BLR UV flux should stay beneath the observed flux, $ \epsilon_{\rm UV} F_{\epsilon_{\rm UV}} < 10^{-11}\,\rm erg\,cm^{-2}\,s^{-1}$ at $\sim 15$~eV. 
The energy density of the BLR emission is $u_{\rm UV}<  \epsilon_{\rm UV} F_{\epsilon_{\rm UV}} d_L^2 /  R'^2 c =  0.2 \,\delta_1^{-2} t_{\rm var,6}^{-1}\,\rm erg\,cm^{-3}$ in the black hole frame. 
 Converting it  to the comoving frame we get 
\beq
f'_{p\gamma} \sim \frac{  4\Gamma^2 u_{\rm UV}}{3 \epsilon_{\rm UV}'}\sigma_{p\gamma}R' <  3.4 \,\left(\frac{u_{\rm UV}}{0.2\,\rm erg\,cm^{-3}}\right)\delta_1^3 t_{\rm var,6}. 
\eeq

Neutrinos carry $\sim 3/4$ of the energy of the charged pions, and about $1/2$ of the $p\gamma$ interactions lead to the production of charged pions. The neutrino flux can thus be estimated by 
\beq\label{eqn:neuFlux}
\left(\epsilon_\nu^2F_{\epsilon_\nu}\right)^{\rm obs} \approx \frac{\delta^4}{4\pi d_L^2} \frac{3}{8} f_{p\gamma}' \gamma_p'^2 m_p c^2 \frac{d \dot{N'}}{d\gamma_p'}\big|_{E_p' \approx 20\,\epsilon_\nu (1+z) / \delta}
\eeq

In the PS scenario, using equations~\ref{eqn:flux_PS}, \ref{eqn:n_gammap} and \ref{eqn:neuFlux} we obtain a peak neutrino flux of 
\bey
\left(\epsilon_\nu^2F_{\epsilon_\nu}\right)^{\rm obs, pk}_{\rm PS} &\sim& \frac{3}{4}\left(\epsilon F_\epsilon\right)_{\rm PS}^{\rm obs, pk}\frac{t'_{\rm PS}(\gamma_{p,b}')}{t'_{\rm lc}}\,f_{p\gamma}' \\ \nonumber
&=& 1.2\times 10^{-11}\, f_{p\gamma}'\delta_1^2\gamma_{p,b,9}'^3 t_{\rm var,6}^{-1} \,\rm erg\,cm^{-2}\,s^{-1}. 
\eey
where $t_{\rm lc}' = R' /c$ is the light crossing time. This is consistent with the observed neutrino flux in the IceCube-170922A event.  

As the proton contribution to $\gamma$-rays is subdominant in the IC scenario, the proton spectrum is poorly constrained. Assuming that $dN'/d\gamma_p'\propto\gamma_p'^{-s}$ and equation~\ref{eqn:Lp_IC} applies, the neutrino flux is estimated to be 
\bey
\left(\epsilon_\nu^2F_{\epsilon_\nu}\right)^{\rm obs, pk}_{\rm IC} &=& \frac{3\,\delta^4}{8}\frac{R'^2}{d_L^2} f_{p\gamma}'u_p'c\left[\frac{(2-s)\gamma_p'^{2-s}}{\gamma_{p,\rm max}'^{2-s} - \gamma_{p,\rm min}'^{2-s}}\right] \\ \nonumber
& \underset{s\sim2.1}=& 5.1\times10^{-10}\,f_{p\gamma}'\delta_1^6\,u_{p,-1}'\,t_{\rm var,6}^2 \gamma_{p,6}'^{2-s}\\ \nonumber
&&\rm erg\,cm^{-2}\,s^{-1}. 
\eey

To summarize, if neutrinos and VHE $\gamma$-rays are produced co-spatially, we expect that the actual neutrino flux level should be much lower than the observed value suggested by \cite{IceCube18a}. But if they are produced in different regions, a strong external photon background would be allowed. The pion production efficiency could be greatly enhanced as discussed in Section~\ref{sec:fpgamma}, leading to an average neutrino flux that is comparable to the IceCube measurement. However, in a multi-zone model, the VHE $\gamma$-ray flare should be delayed compared to the neutrino flare.

\begin{table*}[t!]
\caption{Summary of characteristic polarization and light curve features} \label{table:features}
\centering
\begin{tabular}{c|c|c}
\hline\hline 
Signatures \T & Proton Synchrotron & Inverse Compton \B  \\
\hline
Variability \T  & fast orphan variability in low-E component    & co-variability of both components on short \& long timescales \\
Optical polarization & $\lesssim 10\%$  & $\gtrsim 20\%$\\
$\gamma$-ray polarization &  same as optical & unpolarized \B   \\
\hline
\end{tabular}
\end{table*}

\section{Variability and Polarization Signatures}

The drastic difference in the magnetic field of PS and IC scenarios implies unique features in the light curves and optical polarization. For the PS scenario, the cooling for electrons and protons is dominated by synchrotron, and the cooling time scale is given by
\begin{equation}
\tau'_c=\left({\frac{4}{3}c\sigma_T\frac{u_B}{m_e c^2}(\frac{m_e}{m})^3 \gamma}\right)^{-1}
\end{equation}
Although the size of the emission region is not well constrained in the fitting, typically the PS model has an emission region of $10^{15}-10^{16}~\rm{cm}$ \citep{Boettcher13}. Thus the light crossing time in the comoving frame is $\tau'_{lc}\sim 10^5~\rm{s}$. Our fitting parameters then suggest that the light crossing time scale is between the electron cooling time scale ($\tau'_{e,c}$) and the proton cooling time scale ($\tau'_{p,c}$), i.e.,
\begin{equation}
\tau'_{e,c}<\tau'_{lc}<\tau'_{p,c}
\end{equation}
The fast electron cooling suggests that the optical to soft X-ray light curves should appear symmetric in time, while the slow proton cooling suggests asymmetric patterns from hard X-ray to $\gamma$-ray \citep{Zhang16b}. Additionally, we can expect spike-like optical to soft X-ray flares on top of the active phase due to the fast electron cooling. On the other hand, protons cool slowly by synchrotron, so that the $\gamma$-ray light curve should be rather smooth without any spikes. Therefore, orphan fast variability on top of the active phase in the optical band with smooth $\gamma$-ray light curves can be a signature of the PS scenario. For the IC scenario, since the multi-wavelength emission is from the same primary electron population, the low-energy and high-energy light curves should appear co-variable on both short and long time scales \citep{Chen14}. Based on the multi-wavelength light curves of TXS~0506+056, the optical bands do exhibit some fast variability \citep{IceCube18a}. However, we require better binned {\it Fermi} data to probe PS and IC scenarios by comparing multi-wavelength light curves.

The PS scenario implies a considerably magnetized emission region. The magnetization factor $\sigma$ is defined as the ratio of magnetic energy density over enthalpy. In the PS scenario, it is roughly $L_B/L_{p,kin}$, which is on the order of unity. Numerical simulations have shown that in a magnetized emission region, the optical polarization degree should stay at a low level ($\lesssim 10\%$), and even if a major polarization variability happens, they should quickly revert to the initial level \citep{Zhang16a,Zhang17,Zhang18}. Specifically, magnetic energy dissipation such as a magnetic reconnection triggered by kink instability can show a low polarization degree with nearly constant polarization angle, or the polarization degree drops and reverts to initial value while the polarization angle undergoes a swing. On the other hand, IC scenario implies a weakly magnetized environment, $\sigma\lesssim 0.001$. Under such conditions, the particle acceleration is likely driven by shocks, which can quickly rectify the existing local magnetic field morphology and push the polarization degree to a high level \citep[typically $\gtrsim 20\%$ based on numerical simulations by][]{Laing80,Zhang16a}. Based on the observed spectrum, the optical emission is dominated by primary electron synchrotron without obvious contribution from external thermal photon fields \citep{IceCube18a}. Therefore, the observed $7\%$ polarization degree by Kanata telescope is likely the intrinsic synchrotron polarization, which favors a highly magnetized emission region as in the PS scenario. We suggest that a more detailed analysis of the time-dependent optical polarization degree and angle can better distinguish the PS and IC scenarios. Additionally, based on our fitting parameters and the observed $7\%$ optical polarization degree, the PS scenario predicts a $\sim 7\%$ polarization degree in the $\gamma$-ray \citep{Paliya18}. Due to the slow proton cooling, the $\gamma$-ray polarization degree is generally stationary \citep{Zhang16b}. On the other hand, the IC scenario predicts a $\gamma$-ray polarization degree consistent with zero (upper limit at $\lesssim 3\%$). Future MeV polarimeter such as AMEGO can further constrain the PS and IC scenarios in blazars.

\section{Discussion}

The recent results by \cite{IceCube18a} have opened up new opportunities for observational and theoretical  studies on understanding the origin of high-energy neutrinos as well as the physics of relativistic jets. Below we briefly summarize current theoretical studies and  discuss their similarities and differences compared to our results. \cite{Cerruti18} has performed a thorough parameter survey for one-zone PS and IC scenarios. While both scenarios can produce fittings to the multi-wavelength spectrum, they found that neither contributes adequate neutrino flux compared to the IceCube neutrino flux. \cite{Gao18} have studied a time-dependent one-zone model showing multi-wavelength light curves and neutrino light curves, where they also found inadequate neutrino flux. Both \cite{Keivani18} and \cite{Murase18} have argued against one-zone models based on the X-ray constraints, using the fully numerical time-dependent calculation and analytical approach, respectively. They suggested that to achieve the observed neutrino flux, synchrotron emission of secondary electrons from photomeson processes inevitably overshoot the observed X-ray flux. For the two-zone models, \cite{Murase18} suggested two-zone models that overcome this problem, in which neutrinos are produced in a different region by pp or pgamma interactions. Liu et al. (2018) have also considered such a two-zone model involving both IC + hadronic cascade emissions. \cite{Liu18} have introduced an IC + hadronic cascade model. They suggested that if the emission region locates in a very dense BLR cloud, the $pp$ collision may dominate over photomeson production. This can produce adequate neutrino flux, at the same time the {\it Fermi} $\gamma$-ray emission is a hybrid of IC and hadronic contributions. However, \cite{Murase18} suggests that the strong UV emission from blazars should ionize the BLR clouds along the line of sight, thus a BLR cloud is unlikely to be neutral. Instead, they propose a novel ``neutral beam model'', where the neutrino is beamed while cascade emission is degraded by the de-beaming of secondary pairs. They considered the origin of the external radiation field, and argued that it could be provided by the sheath region of the jet. \cite{MAGIC18} have also put forward a spine-sheath jet model where the emission region is composed of a fast-moving spine and a slow-moving sheath in the emission region. In this paper, we put forward a different two-zone model, where the energetic particles propagate from an optically-thick to an optically-thin environment for VHE $\gamma$-rays. We find that this model can also produce the observed neutrino flux and the multi-wavelength spectrum. We also investigate generic variability and polarization signatures of the PS and IC scenarios. These features only depend on the magnetic field strength of the emission region, which is intrinsic to the models regardless of the specific parameters or the co-spatiality of the neutrino and multi-wavelength emission. Therefore, our conclusions apply to a much wider parameter regime than what we have investigated in this paper. However, a self-consistent time-dependent study of the multi-wavelength light curves and polarization patterns is far beyond the scope of this paper, which will be detailed in a future study. 

Our main conclusions are:
\begin{enumerate}
\item The PS scenario may exhibit fast variability in the low-energy spectral component, but no fast variability counterpart in the high-energy component. It predicts low optical polarization degree throughout the active state, and a $\gamma$-ray polarization degree at the same level as the optical band.
\item The IC scenario exhibits co-variability of low- and high-energy spectral components on both short and long time scales. It should have a highly variable optical polarization degree that can reach $\gtrsim 20\%$, whereas its $\gamma$-ray emission is nearly unpolarized.
\item  In a simple one-zone model, the average neutrino flux is $\sim1\%$ of the observed level. In a two-zone model and when there is an intense external UV field, the neutrino flux could be higher. In the latter case, the neutrino detection comes before the $100~\rm{GeV}$ flare for both two-zone PS and IC scenarios.
\end{enumerate}
We summarize in Table~\ref{table:features} the generic variability and polarization features for readers' reference.

\acknowledgments{We thank Kohta Murase for helpful comments. We acknowledge helpful discussions with B. Dingus. HZ acknowledge helpful comments from Dimitrios Giannios. HZ acknowledges support from Fermi Guest Investigator program Cycle 10, grant number 80NSSC17K0753. KF acknowledges the support of a Joint Space-Science Institute prize postdoctoral fellowship at the University of Maryland. HL acknowledges the support by the LANL/LDRD program and the DoE/OFES program.}

\clearpage

\end{document}